\documentclass{article}

\usepackage{arxiv}

\usepackage[utf8]{inputenc} 
\usepackage[T1]{fontenc}    
\usepackage{hyperref}       
\usepackage{url}            
\usepackage{booktabs}       
\usepackage{amsfonts}       
\usepackage{nicefrac}       
\usepackage{microtype}      
\usepackage{lipsum}		
\usepackage{graphicx}
\usepackage{natbib}
\usepackage{doi}
\usepackage{amsmath} 
\usepackage{amssymb}
\usepackage{xcolor}

\title{Robust Ferrimagnetic Ground State and Suppressed Superconductivity in Two-Dimensional HC$_6$}

\author{ \href{https://orcid.org/0009-0004-2196-8245}{\includegraphics[scale=0.06]{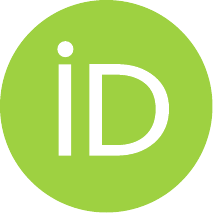}\hspace{1mm}Jakkapat Seeyangnok} \\
	Department of Physics\\
    Faculty of Science\\
	Chulalongkorn University\\
	Bangkok, Thailand \\
	\texttt{jakkapatjtp@gmail.com} \\
	\And
	\href{https://orcid.org/0000-0002-8450-7751}{\includegraphics[scale=0.06]{orcid.pdf}\hspace{1mm}Udomsilp Pinsook} \\
	Department of Physics\\
    Faculty of Science\\
	Chulalongkorn University\\
	Bangkok, Thailand \\
	\texttt{Udomsilp.P@Chula.ac.th} \\
}

\begin{document}
\maketitle

\begin{abstract}
Two-dimensional hydrogenated graphene (HC$_6$) represents a promising platform for exploring emergent electronic phases. Owing to its high electronic density of states at the Fermi level, HC$_6$ is expected to support phonon-mediated superconductivity, with a calculated critical temperature ($T_c$) of 37.4 K in the paramagnetic metallic phase. However, spin-polarized first-principles calculations reveal that HC$_6$ stabilizes in a ferrimagnetic ground state, which is energetically favored by 0.175 eV per unit cell over the paramagnetic metallic phase. This large energy difference significantly exceeds $k_B T$ at room temperature, indicating robust magnetic order. Although the superconducting condensation energy lowers the total energy by ~7 meV, the superconducting phase remains metastable. These results highlight the dominant role of magnetism in HC$_6$ and illustrate how a high electronic density of states can drive competing instabilities in hydrogenated two-dimensional materials, offering design principles for carbon-based magnetic systems.
\end{abstract}


\section{Introduction}

Two-dimensional (2D) materials have emerged as versatile platforms for exploring quantum phenomena and designing novel electronic phases. Since the discovery of graphene in 2004 \cite{novoselov2004electric}, the field has expanded dramatically, uncovering systems that host magnetism, superconductivity, and topological order.

Beyond their structural flexibility, 2D materials enable the study of magnetism at the atomic scale. Reduced dimensionality gives rise to diverse ground states such as ferromagnetism (FM), antiferromagnetism (AFM), and half-metallicity (HM), which are highly attractive for spintronic and valleytronic applications \cite{MoS2-2_2012,valley2016,MoS2_HM_2015,MoS2_2016,MoS2_HM_2017,MoS2_HM_2018,MoS2_AFM_2021}. Transition metal dichalcogenide (TMD) monolayers exemplify these behaviors: VS$_2$ has been proposed as an intrinsic FM material \cite{VS2_2013,VS2_2017}, while compounds such as CoCl$_2$, CoBr$_2$ \cite{CoCl2_2018,CoBr2_2019,CoCl2_2023}, MnS$_2$, and MnSe$_2$ \cite{MnS2_2014} display semiconducting magnetic states. More recently, Janus monolayers, created by selective atomic substitution that breaks mirror symmetry, have provided new routes to engineer magnetic and electronic properties. Examples include MoSSe \cite{trivedi2020room,lu2017janus}, WSSe \cite{trivedi2020room}, PtSSe \cite{sant2020synthesis}, and MnSSe \cite{mnsse2022prb}. Janus transition metal chalcogenide hydrides have also attracted attention, exhibiting half-metallic ferromagnetism and emerging as promising spintronic candidates \cite{1tcrsh2024,crsh_earth2025,seeyangnok2025competition}.

Alongside magnetism, superconductivity in 2D materials has become a major focus. Numerous studies have shown that low-dimensional systems can sustain superconducting phases when the density of states at the Fermi level and electron–phonon coupling are favorable \cite{gao2017prediction,bekaert2017evolution,zhao2019two,zhao2020mgb,gao2021undamped,singh2022high,sevik2022high,dong2022superconductivity,wines2023high}. While pristine graphene is not superconducting, modifications such as hydrogenation and strain have been proposed to induce it \cite{profeta2012phonon}. For instance, graphane has been predicted to reach $T_c$ values above 90 K \cite{sofo2007graphane,savini2010first}, while twisted bilayer graphene hosts unconventional superconductivity near the magic angle \cite{cao2018unconventional}. Progress has extended to hydrogen-rich 2D systems, where higher $T_c$ values may be realized under ambient or near-ambient conditions. Examples include hydrogenated MgB$_2$, with $T_c$ up to 100 K under strain \cite{bekaert2019hydrogen}, and hydrogenated phosphorus carbide (HPC$_3$), which shows $T_c$ enhancement from 31 K to 57.3 K under tensile strain \cite{li2022phonon}. Similar improvements have been reported in other hydrogenated 2D materials \cite{seeyangnok2024super,seeyangnok2024superconductivity,pinsook2025superconductivity,ul2024superconductivity,yan2022enhanced,han2023theoretical,xue2024realization,seeyangnok2025high_npj2d,seeyangnok2025hydrogenation_nanoscale}.

Here, we investigate the electronic, magnetic, and superconducting properties of hydrogenated graphene HC$_6$ using first-principles calculations. We find that although the high density of states near the Fermi level suggests the possibility of phonon-mediated superconductivity, HC$_6$ stabilizes in a ferrimagnetic ground state that is energetically favored over the non-magnetic metallic phase. This robust magnetic ordering renders superconductivity metastable, though external factors such as strain or doping may alter the balance. Our results highlight HC$_6$ as a platform for studying the interplay between magnetism and superconductivity in hydrogenated carbon-based 2D systems, providing insight into phase competition and potential routes for functional tuning.

\section{Computational Details}

All calculations were performed within density functional theory (DFT) using the \textsc{Quantum Espresso} package \cite{giannozzi2009quantum,giannozzi2017advanced}, with initial crystal structures generated in VESTA \cite{momma2011vesta}. Structural optimization was carried out using the BFGS algorithm \cite{BFGS,liu1989limited}, with full relaxation until forces were below $10^{-5}$ eV/\AA. A vacuum spacing of 20 \AA{} was applied to avoid spurious interlayer interactions. The generalized gradient approximation with the PBE functional \cite{perdew1996generalized} and optimized norm-conserving Vanderbilt pseudopotentials \cite{hamann2013optimized,schlipf2015optimization} were employed. Plane-wave cutoffs of 80 Ry (wavefunctions) and 320 Ry (charge density) were used. Brillouin zone integrations employed a $16 \times 16 \times 1$ Monkhorst–Pack $k$-point mesh for the primitive cell and $8 \times 8 \times 1$ for $2 \times 2 \times 1$ supercells of HC$_6$, with a Methfessel–Paxton smearing of 0.02 Ry \cite{methfessel1989high}.  

For superconductivity, electron–phonon coupling (EPC) was calculated using the Wannier–Fourier interpolation method \cite{giustino2017electron,giustino2007electron} as implemented in the EPW package \cite{noffsinger2010epw,ponce2016epw}. The anisotropic Migdal–Eliashberg equations \cite{margine2013anisotropic,pinsook2024analytic} were solved self-consistently on the imaginary axis to obtain the superconducting gap and critical temperature ($T_c$). Calculations employed dense meshes of $80 \times 80 \times 1$ $k$-points and $40 \times 40 \times 1$ $q$-points, with convergence checked against finer grids. A Gaussian broadening of 0.10 eV (electrons) and 0.50 meV (phonons) was applied to Dirac delta functions. The Coulomb pseudopotential was set to $\mu^* = 0.10$, a commonly adopted value for 2D superconductors. The Fermi surface thickness and Matsubara cutoff energies were chosen to ensure numerical convergence.  

\section{Phase Properties}
    \begin{figure}[tbh!]
        \centering
		\includegraphics[width=12cm]{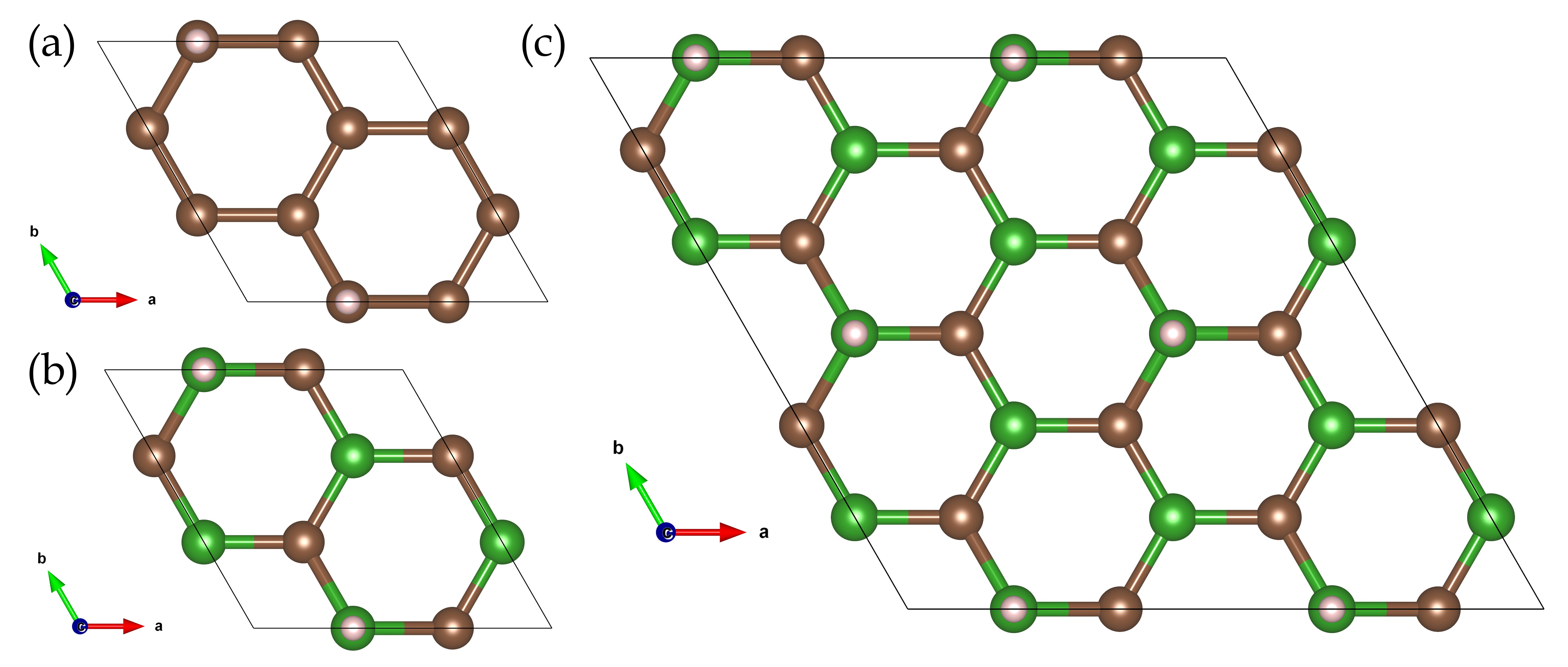}
		\caption{Structural representation of HC6 in different magnetic configurations. (a) The normal metallic state with either non-magnetic or ferromagnetic ordering, in which all carbon atoms are represented by brown spheres and hydrogen atoms by pink spheres.(b) Magnetic ordering is introduced, where the brown and green spheres represent carbon atoms with opposite spin orientations (either up and down or down and up). Hydrogen atoms remain pink. (c) A 2$\times$2$\times$1 supercell of HC6 is constructed to explore other possible magnetic arrangements, showing extended periodic magnetic ordering with alternating spin orientations. The coordinate axes (a, b, c) indicate the lattice directions.  }
		\label{fig:HC6_magnetic}
    \end{figure}
To explore the magnetic ordering in HC$_6$, spin-polarized density functional theory (DFT) calculations were performed for several configurations, including ferromagnetism (FM), G-type antiferromagnetism (GAF), C-type antiferromagnetism (CAF), and the non-magnetic (NM) metallic state. Total energy comparisons for these phases are summarized in Fig.~\ref{fig:HC6_magnetic}(a,b). Calculations were carried out for both the primitive cell and a $2 \times 2 \times 1$ supercell, yielding consistent results.  

All initial magnetic configurations converged to the same solution: a ferrimagnetic phase. In this ordering, carbon atoms on different sublattices carry antiparallel magnetic moments of unequal magnitude, producing a finite net magnetization. The ferrimagnetic phase is energetically favored by 0.175 eV per unit cell compared to the NM metallic state, establishing it as the robust ground state of HC$_6$. This large stabilization energy, well above thermal fluctuations at room temperature, indicates strong magnetic ordering.  

Given the high density of states at the Fermi level, HC$_6$ could in principle support phonon-mediated superconductivity. However, as discussed later, the superconducting condensation energy ($\sim$7 meV) is insufficient to overcome the stability of the ferrimagnetic phase, leaving superconductivity metastable relative to the true ground state.

\section{Electronic Structures}
    \begin{figure}[tbh!]
        \centering
		\includegraphics[width=14cm]{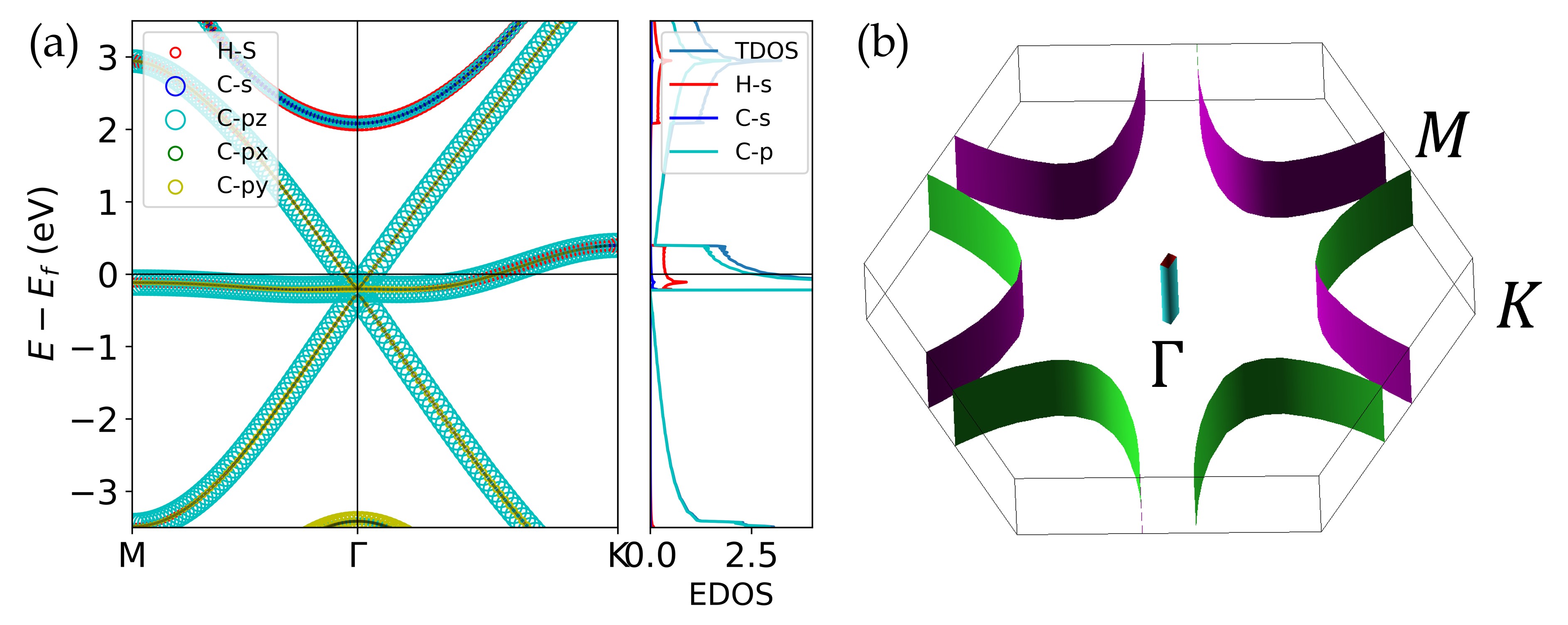}
		\caption{(a) Orbital-resolved electronic band structure showing the contribution of different atomic orbitals. The color-coded markers represent orbital character: red for H-$s$, blue for C-$s$, light blue for C-$p_z$, green for C-$p_x$, and yellow for C-$p_y$ orbitals. The energy is plotted relative to the Fermi level \( E_F \) along high-symmetry points \( M-\Gamma-K \). The corresponding total and partial density of states (TDOS) is shown on the right. (b) Fermi surface of the material, illustrating the topology and symmetry around the \( \Gamma \) point. The green and purple surfaces highlight distinct segments of the Fermi surface, with high-symmetry points \( M \) and \( K \) labeled.}
		\label{fig:orbital_fermi}
    \end{figure}
A high density of states (DOS) near the Fermi level is often linked to instabilities in metallic systems, which may drive structural distortions, superconductivity, or magnetic ordering \cite{ackland2004origin}. The calculated electronic band structure and projected DOS of HC$_6$ confirm metallic behavior, with multiple band crossings at the Fermi level and a pronounced DOS peak at $E_F$. Two main bands contribute strongly to this feature, suggesting enhanced electron–phonon coupling. Orbital projections reveal that both carbon $p_z$ and hydrogen $s$ orbitals dominate the electronic states around $E_F$, indicating that both species are crucial to the low-energy physics.  

This behavior is consistent with earlier findings in partially hydrogenated graphene, where nearly flat bands and a high DOS near the Fermi level promote magnetic ordering \cite{lu2016ferromagnetism}. Similar trends have also been reported in Janus $MX$H monolayers ($M = \text{Ti, Zr, Hf}$; $X = \text{S, Se, Te}$), in which a high DOS near $E_F$ drives competition between superconductivity and ferromagnetism, with the latter generally favored \cite{seeyangnok2025competition}. In contrast, other Janus monolayer hydrides, such as WSH and MoSH, remain non-magnetic metals and undergo a transition to superconductivity below their critical temperature \cite{liu2022two,ku2023ab,seeyangnok2024superconductivity,seeyangnok2024super}.  

These comparisons emphasize that the electronic structure of HC$_6$, with its high DOS at the Fermi level, provides a natural stage for competing instabilities. This motivates a detailed investigation of both magnetic and superconducting phases, as discussed in the following sections.

\subsection{Metallic Phase: Electronics and Superconductivity}
    \begin{figure}[ht]
        \centering
		\includegraphics[width=13cm]{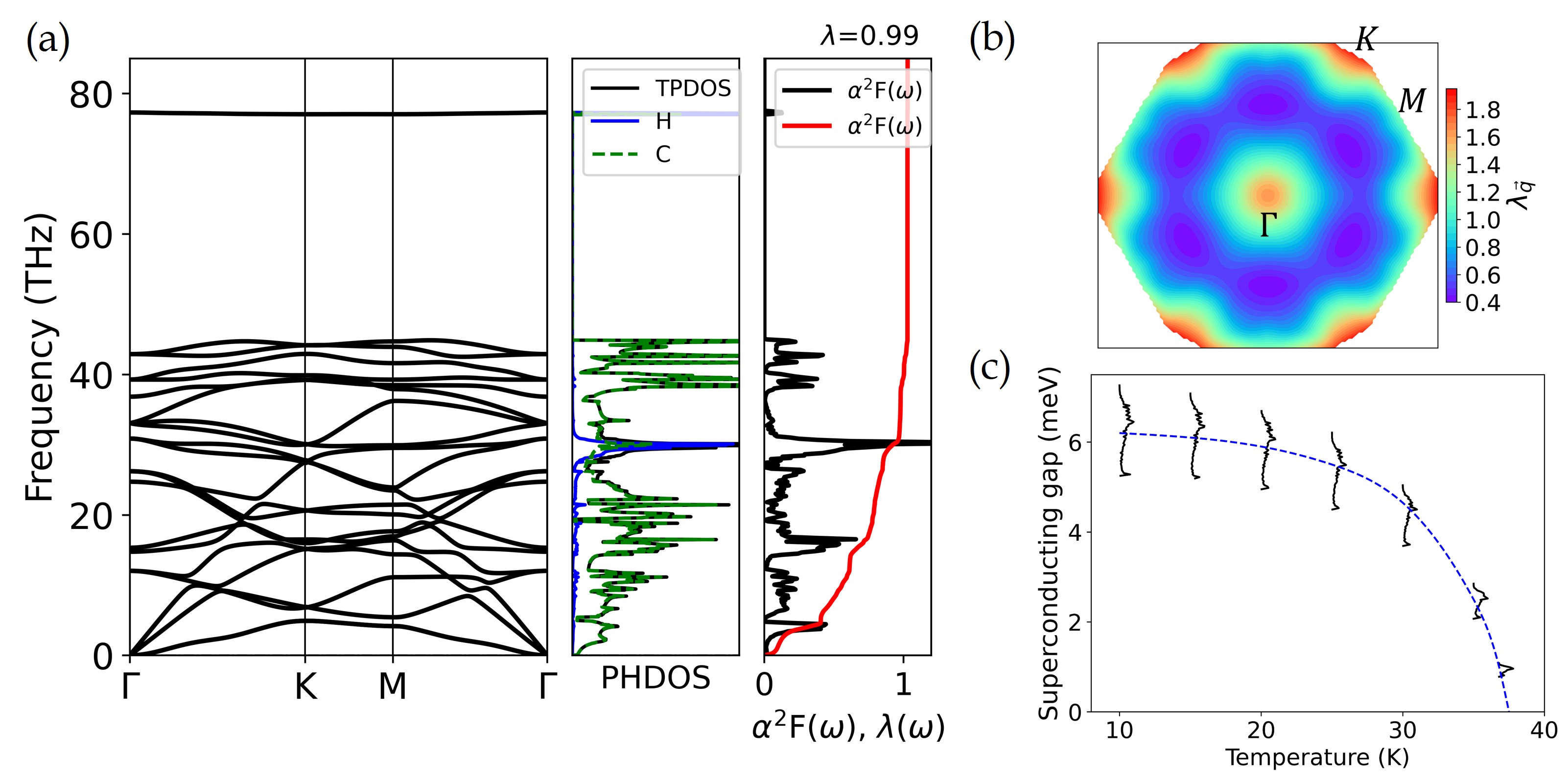}
    \caption{
    (a) Phonon dispersion along high-symmetry points \( \Gamma-K-M-\Gamma \) with the corresponding projected phonon density of states (PHDOS) for H (blue) and C (green). The Eliashberg spectral function \( \alpha^2F(\omega) \) (red) and the cumulative electron-phonon coupling parameter \( \lambda(\omega) \) (black) are shown on the right. The total phonon density of states (TPDOS) is displayed in black.  (b) Electron-phonon coupling density \( \lambda_q \) plotted on the two-dimensional Brillouin zone, indicating stronger coupling regions in red and weaker regions in blue. (c) Superconducting gap as a function of temperature, showing a decreasing trend with increasing temperature. The dashed blue curve represents the fitting to the Bardeen-Cooper-Schrieffer (BCS) theory.  
    }
    \label{fig:phonon_sc}
    \end{figure}
Figure~\ref{fig:phonon_sc}(b) presents the momentum-resolved electron–phonon coupling strength, $\lambda_{\boldsymbol{q}}$, across the Brillouin zone. The coupling is highly anisotropic: the strongest interactions ($\lambda_{\boldsymbol{q}} > 1.8$) occur near the $K$ point and zone edges, while the central region around $\Gamma$ remains relatively weak ($\lambda_{\boldsymbol{q}} < 1.0$). Complementary information is provided by the Eliashberg spectral function $\alpha^2F(\omega)$ and its integral $\lambda(\omega)$, shown in Fig.~\ref{fig:phonon_sc}(a). The spectral function exhibits multiple peaks, notably from the acoustic ZA mode near 16 meV and optical modes around 65 meV and 118 meV. The high-frequency peak at 118 meV originates primarily from hydrogen vibrations. The integrated coupling saturates at $\lambda = 0.99$, confirming strong electron–phonon interactions.  

The superconducting gap $\Delta_k$ on the Fermi surface was obtained by solving the anisotropic Migdal–Eliashberg equations. As shown in Fig.~\ref{fig:phonon_sc}(c), the gap reaches $\Delta_0 \approx 6.2$ meV at low temperature ($T \approx 0$ K) and closes at $T_c \approx 37.4$ K. The ratio $2\Delta_0/k_B T_c = 3.85$ exceeds the BCS value of 3.53, consistent with strong-coupling superconductivity. The condensation energy lowers the total energy of the metallic phase by about 7 meV at 10 K, indicating that the superconducting state is stabilized relative to the normal metal.  

Nevertheless, as discussed below, this stabilization is far smaller than the 0.175 eV energy gain associated with ferrimagnetic ordering, confirming that superconductivity in HC$_6$ is only a metastable phase.

\subsection{Ferrimagnetic Phase: Electronics and Spin Polarization}
    \begin{figure}[h!]
    \centering
	\includegraphics[width=16cm]{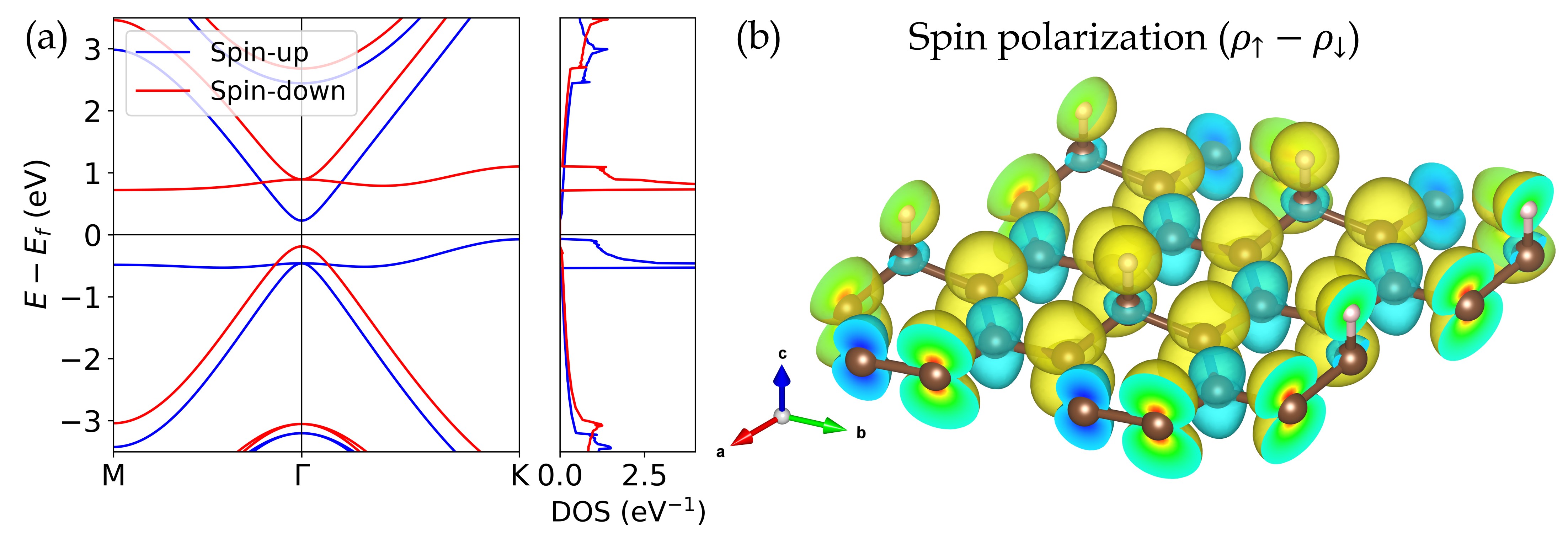}
    \caption{
    (a) Spin-polarization resolved electronic band structure along high-symmetry points \( M-\Gamma-K \). The spin-up and spin-down channels are represented by blue and red lines, respectively. The corresponding density of states (DOS) is shown on the right. (b) Spin-resolved charge density with an isosurface value of 0.004 e/Å\(^3\), illustrating the spatial distribution of spin polarization (\(\rho_\uparrow - \rho_\downarrow\)). The blue and yellow regions indicate spin-up and spin-down densities, respectively. The crystallographic axes \( a \), \( b \), and \( c \) are denoted by the arrows.  
    }
    \label{fig:spin_polarization}
    \end{figure}
The ferrimagnetic ground state is evidenced by both the spin-resolved electronic structure and the real-space spin density distribution (Fig.~\ref{fig:spin_polarization}). The spin-polarized band structure and density of states (DOS) [Fig.~\ref{fig:spin_polarization}(a)] exhibit a pronounced exchange splitting between the spin-up (blue) and spin-down (red) channels in the vicinity of the Fermi level, confirming the presence of long-range magnetic order. Notably, an energy gap of approximately 0.42~eV opens at the $\Gamma$ point, reflecting the exchange-induced band reconstruction. The asymmetry in the DOS at $E_F$ further highlights the net spin polarization, in agreement with a ferrimagnetic configuration.

The corresponding real-space spin density, shown in Fig.~\ref{fig:spin_polarization}(b), illustrates the alternating spin polarization on inequivalent carbon sites. Yellow lobes mark regions with an excess of spin-up electrons, while blue lobes indicate spin-down excess. The unequal magnitude of local moments on opposite sublattices yields a ferrimagnetic configuration with a finite net magnetization.  

Quantitatively, the system exhibits a total magnetization of 1.02~$\mu_B$/cell. These results demonstrate that HC$_6$ stabilizes in a ferrimagnetic ground state, where partially compensated spin moments coexist with metallic electronic character. This ferrimagnetic ordering is energetically favored over the non-magnetic metallic and superconducting phases, confirming its role as the true ground state of the system.

\begin{figure}[ht]
        \centering
		\includegraphics[width=8cm]{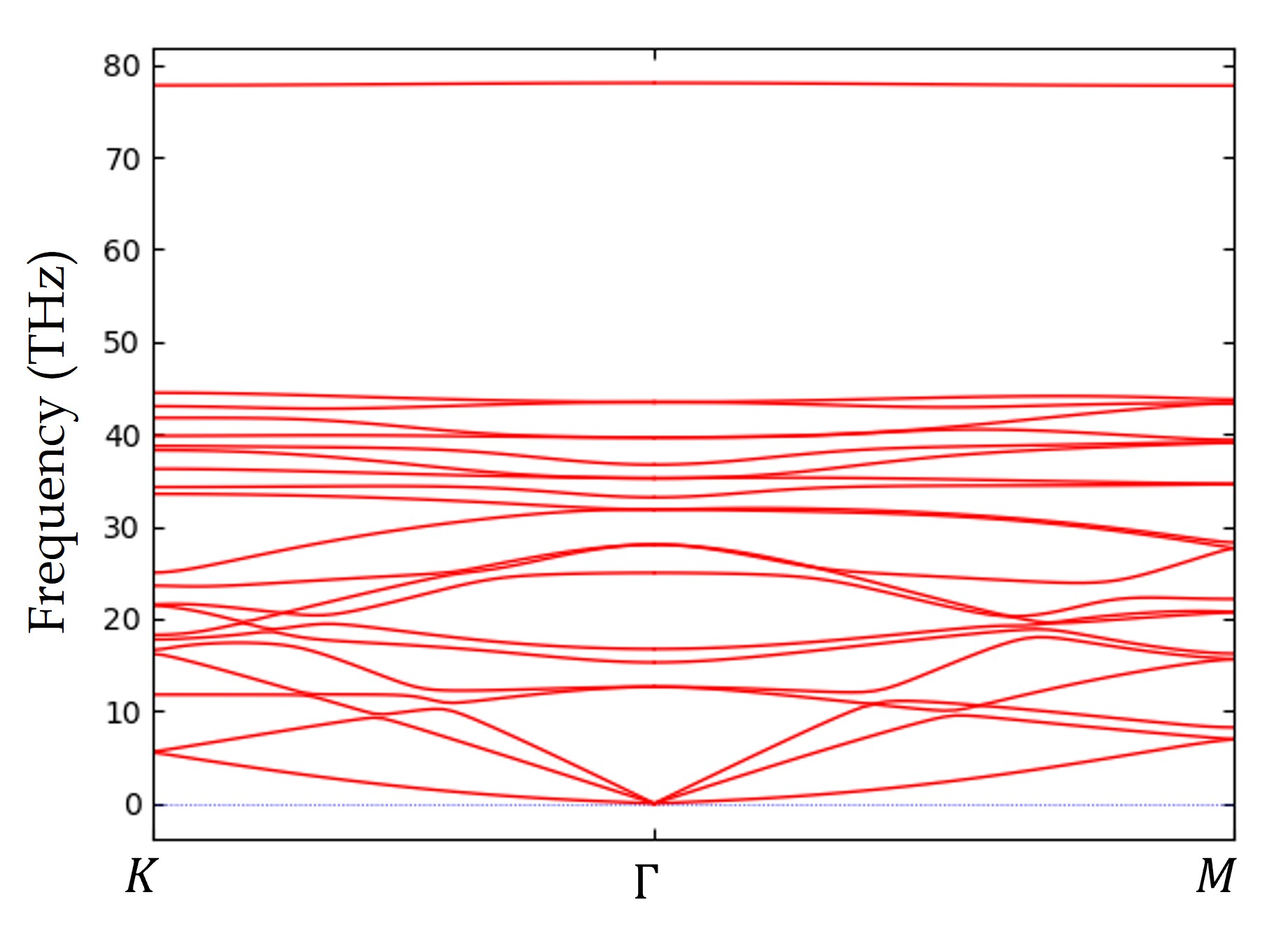}
    \caption{
    Phonon dispersion computed using the PHONOPY package with a \(2 \times 2 \times 1\) supercell incorporating magnetic ordering. The absence of imaginary frequencies across the Brillouin zone indicates dynamically stable phonon modes.
    }
    \label{fig:phonon_dispersion}
    \end{figure}

To assess the dynamical stability of the magnetic ground state, we computed the phonon dispersion of the ferrimagnetic phase using the finite displacement method implemented in Phonopy. The phonon spectrum along the $K$–$\Gamma$–$M$ path (Fig.~\ref{fig:phonon_dispersion}) shows no imaginary frequencies, confirming that HC$_6$ is dynamically stable in its ferrimagnetic configuration. This result validates the robustness of the predicted magnetic ground state and supports the reliability of the associated electronic and magnetic properties.

\section{Competition between Superconductivity and Ferrimagnetism}

Superconductivity and magnetism are intrinsically antagonistic: conventional phonon-mediated superconductivity relies on spin-singlet Cooper pairs, whereas magnetic order breaks time-reversal symmetry and introduces pair-breaking fields. In HC$_6$, this conflict manifests as a competition between ferrimagnetism and superconductivity.  

Our first-principles calculations show that the ferrimagnetic phase is strongly stabilized, with a total energy lower by 175 meV per cell compared to the non-magnetic metallic state. In contrast, when superconductivity develops in the metallic phase below $T_c = 37.4$ K, the condensation energy lowers the total energy by only $\sim$7 meV. This disparity demonstrates that the magnetic exchange splitting overwhelms the superconducting pairing energy, rendering the superconducting state metastable under equilibrium conditions.  

Thus, while HC$_6$ possesses the electronic features required for strong-coupling superconductivity, robust ferrimagnetism dominates its ground-state properties. These findings highlight HC$_6$ as a model two-dimensional system where high density of states at the Fermi level drives competing instabilities, and where external control parameters such as strain, doping, or gating may tip the balance between superconductivity and magnetism.

\section{Conclusion}

We have explored the competing electronic instabilities of HC$_6$ using first-principles calculations. Spin-polarized calculations reveal that the ferrimagnetic phase is the true ground state, stabilized by 175 meV per cell relative to the non-magnetic metallic configuration. This phase arises from unequal antiparallel spin moments on different sublattices, yielding a net magnetization.  

In contrast, the non-magnetic metallic phase supports strong electron–phonon coupling that drives a superconducting instability with a critical temperature of 37.4 K. The associated condensation energy lowers the metallic state by only $\sim$7 meV, which is insufficient to overcome the energetic stabilization of ferrimagnetism. Consequently, superconductivity remains metastable under equilibrium conditions.  

These findings demonstrate that HC$_6$ is a prototypical two-dimensional system where robust magnetism suppresses a latent superconducting state. External control parameters such as carrier doping, epitaxial strain, or applied pressure may provide routes to tune this balance, potentially stabilizing superconductivity or enabling coexistence with magnetic order. Such control would open avenues for novel quantum phases and functional devices in the realms of spintronics and magnetic superconductors.


    \section*{Data Availability}
    The data that support the findings of this study are available from the corresponding
    authors upon reasonable request.
    
    \section*{Code Availability}
    The first-principles DFT calculations were performed using the open-source Quantum ESPRESSO package, available at \url{https://www.quantum-espresso.org}, along with pseudopotentials from the Quantum ESPRESSO pseudopotential library at \url{https://pseudopotentials.quantum-espresso.org/}. Electron-phonon coupling and related properties were computed using the EPW code, available at \url{https://epw-code.org/}.

    \section*{Acknowledgements}
	This research project is supported by the Second Century Fund (C2F), Chulalongkorn University. We acknowledge the supporting computing infrastructure provided by NSTDA, CU, CUAASC, NSRF via PMUB [B05F650021, B37G660013] (Thailand). URL:www.e-science.in.th.

    \section*{Author Contributions}
    Jakkapat Seeyangnok performed all of the calculations, analysed the results, wrote the first draft manuscript, and coordinated the project. Udomsilp Pinsook analysed the results and wrote the manuscript.

    \section*{Conflict  of Interests}
    The authors declare no competing financial or non-financial interests.

	\section*{References}
    \bibliographystyle{unsrt} 
	\bibliography{references}

\end{document}